# Thermal generation, manipulation and detection of skyrmions

Zidong Wang[1,2,*], Minghua Guo[1,3,*], Heng-An Zhou[1,2,*], Le Zhao[1,2], Teng Xu[1,2], Riccardo Tomasello[4], Hao Bai[1,2], Yiqing Dong[1,2], Soong Geun Je[5,6], Weilun Chao[5], Hee-Sung Han[7], Suseok Lee[7], Ki-Suk Lee[7,8], Yunyan Yao[9], Wei Han[9], Cheng Song[10], Huaqiang Wu[3], Mario Carpentieri[11], Giovanni Finocchio[12], Mi-Young Im[5,7], Shi-Zeng Lin[13,†] and Wanjun Jiang[1,2,†]

[1]*State Key Laboratory of Low-Dimensional Quantum Physics and Department of Physics, Tsinghua University, Beijing 100084, China*

[2]*Frontier Science Center for Quantum Information, Tsinghua University, Beijing 100084, China*

[3]*Institute of Microelectronics, Tsinghua University, Beijing 100084, China*

[4]*Institute of Applied and Computational Mathematics, FORTH, GR-70013, Heraklion-Crete, Greece*

[5]*Center for X-ray Optics, Advanced Light Source, Lawrence Berkeley National Laboratory, Cyclotron Road, Berkeley, CA 94720, USA*

[6]*Department of Physics, Chonnam National University, Gwangju 61186, Korea*

[7]*School of Materials Science and Engineering, Ulsan National Institute of Science and Technology, Ulsan 44919, Korea*

[8]*Daegu Gyeongbuk Institute of Science and Technology, Daegu 711-873, Korea*

[9]*International Center for Quantum Materials, School of Physics, Peking University, Beijing, 100871, China*

[10]*Key Laboratory of Advanced Materials (MOE), School of Materials Science and Engineering, Tsinghua University, Beijing 100084, China*

[11]*Department of Electrical and Information Engineering, Polytechnic of Bari, Bari 70125, Italy*

[12]*Department of Mathematical and Computer Sciences, Physical Sciences and Earth Sciences, University of Messina, Messina 98166, Italy*

[13]*Theoretical Division, T-4 and CNLS, Los Alamos National Laboratory, Los Alamos, New Mexico 87545, USA*

[*] These authors contributed equally to this work.

[†] To whom correspondence should be addressed:

jiang_lab@tsinghua.edu.cn, szl@lanl.gov.

**Recent years have witnessed significant progresses in realizing skyrmions in chiral magnets[1-4] and asymmetric magnetic multilayers[5-13], as well as their electrical manipulation[2,7,8,10]. Equally important, thermal generation, manipulation and detection of skyrmions can be exploited for prototypical new architecture with integrated computation[14] and energy harvesting[15]. It has yet to verify if skyrmions can be purely generated by heating[16,17], and if their resultant direction of motion driven by temperature gradients follows the diffusion or, oppositely, the magnonic spin torque[17-21]. Here, we address these important issues in microstructured devices made of multilayers – [Ta/CoFeB/MgO]$_{15}$, [Pt/CoFeB/MgO/Ta]$_{15}$ and [Pt/Co/Ta]$_{15}$ integrated with on-chip heaters, by using a full-field soft X-ray microscopy. The thermal generation of densely packed skyrmions is attributed to the low energy barrier at the device edge, together with the thermally induced morphological transition from stripe domains to skyrmions. The unidirectional diffusion of skyrmions from the hot region towards the cold region is experimentally observed. It can be theoretically explained by the combined contribution from repulsive forces between skyrmions, and thermal spin-orbit torques in competing with magnonic spin torques[17,18,20,21] and entropic forces[22,]. These thermally generated skyrmions can be further electrically detected by measuring the accompanied anomalous Nernst voltages[23]. The on-chip thermoelectric generation, manipulation and detection of skyrmions could open another exciting avenue for enabling skyrmionics, and promote interdisciplinary studies among spin caloritronics[15], magnonics[24] and skyrmionics[3,4,12].**

The nontrivial spin topology and particle-like behaviors of skyrmions have recently stimulated immense interests in the spintronics community[4,12,25]. For practical applications, efficient generation, manipulation and detection of skyrmions are essential. On one side, electrical currents and/or current-induced spin-orbit torques (SOTs) are utilized in various metallic systems[2,8,10]. On the other side, a pure thermal generation, manipulation and detection of skyrmions have not been studied experimentally, which motivate the present study. Such experiments, while remaining challenge in bulk samples and large-area films, can be done in microstructured devices integrated with on-chip heaters, as successfully demonstrated in the present study. Namely, elevating local temperatures facilitates the skyrmion formation out of other competing magnetic phases by overcoming the corresponding energy barriers[16,17]. Our results can be particularly useful for studying skyrmions and their dynamics in magnetic metals and insulators[26,27], and may lead to various technologically relevant physics/device concepts to be investigated in unconventional computing[14].

**Experimental demonstration of thermal generation of skyrmions.**

The asymmetric multilayers used in the present study are: $[Ta/Co_{20}Fe_{60}B_{20}/MgO]_{15}$, $[Pt/Co_{60}Fe_{20}B_{20}/MgO/Ta]_{15}$ and $[Pt/Co/Ta]_{15}$, which were typified with a monotonic increase of the interfacial Dzyaloshinskii-Moriya interaction (DMI) strengths[8,10,13,28] and damping paramters[10]. Magnetometry measurements reveal typical hysteresis loops that are reminiscence of multilayers hosting either Néel-type[10,11,13] or hybrid-type skyrmions[29,30]. Since the size of skyrmions in multilayers are generally smaller than 200 nm, we employed a soft X-ray full-field transmission microscope (XM-1) with a spatial resolution down to 20 nm at the beamline 6.1.2 at the Advanced Light Souce of Lawrence Berkeley National Laboratory[10]. Rich sample environment of XM-1 enables the dynamics of skyrmions stimulated by perpendicular magnetic field ($\mu_0 H_\perp$), electrical current ($j_e$), temperature ($T$) and temperature gradient ($\Delta T(x)$) to be investigated. Magnetic imaging has been conducted at the Fe $L_3$ edge (708.5 eV) in the $[Ta/Co_{20}Fe_{60}B_{20}/MgO]_{15}$, and at the Co $L_3$ edge (778.5 eV) in the $[Pt/Co_{60}Fe_{20}B_{20}/MgO/Ta]_{15}$ and the $[Pt/Co/Ta]_{15}$ multilayers. Note that the dominant features of the thermal generation of skyrmions in these three multilayers remain the same, indicating the phenomenology revealed in our experiments are generic in multilayers hosting skyrmion. We focus on the $[Ta/Co_{20}Fe_{60}B_{20}/MgO]_{15}$ multilayers for which exhibit the strongest X-ray magnetic circular dichroism (XMCD) contrast and it is a low pinning stack[13,14], unless otherwise specified.

We have fabricated devices by integrating magnetic multilayers with on-chip heaters and thermometers made of Ta (20 nm)/Pt (50 nm) for both *in-situ* control/measure of temperatures and anomalous Nernst measurements. Those devices were prepared onto the insulating $Si_3N_4$ (100 nm thick) membranes, in order to ensure X-ray transmission imaging. An optical image of the device is shown in Fig. 1A. By applying a pulse voltage ($V_h$) into the Ta/Pt heater, a local temperature gradient $\Delta T(x)$, is created through heat dissipation in the underneath insulating $Si_3N_4$ thin layer. The dissipated heat arrives at the multilayer that could generate skyrmions from different background orderings. Furthermore, the accompaned temperature gradient produces diffusion of skyrmions, and allows an electrical detection of thermally generated skyrmions via measuring the anomalous Nernst voltage ($V_{ANE}$) between two contacts ($C_{1,2}$). Additionally, symmetric heaters located on both sides of multilayer enable the directional control of skyrmion generation to be done. Calibrations of temperatures and temperature gradients, discussed in Part 8 of Supplementary Materials, suggest the existence of quasilinear temperature gradients in the multilayer. For the maximum voltage applied into the Ta/Pt heater, the Oersted field at the multilayer was calculated to be less than 0.22 mT, thus its influence on skyrmion dynamics is negligible. The threshold depinning electrical current at room temperature (below which skyrmions do not move), is estimated to be of the order of $J_{th} \approx 10^5$ A/cm$^2$ by applying electrical currents to the multilayers, which is comparable with similar multilayers[13].

The temperature profile of the whole device on the 100 nm-thick Si$_3$N$_4$ membrane (500 × 500 μm$^2$) was computed by the COMSOL Multiphysics software with the material specific parameters, as shown in Fig. 1B. Detailed descriptions are given in Method section and Supplementary Materials. A linescan of the device (cyan line) is shown in the lower panel of Fig. 1B, which confirms a varying temperature profile along the *x*-axis. For different voltages $V_h$ applied to the upper Heater (H1), temperatures of both heaters ($T^S_{H1,2}$) and the upper/lower edges of the multilayer ($T^S_{U,L-E}$) were simulated and shown in the right panel of Fig. 1B. From the temperature difference between $T^S_{U-E}$ and $T^S_{L-E}$, a linear temperature gradient $\Delta T(x)$ along the *x*-axis in the multilayer can be found. The experimentally measured temperatures of both heaters (H1,2) agree with the COMSOL simulations.

Representative images of the thermal generation of skyrmions in the [Ta/Co$_{20}$Fe$_{60}$B$_{20}$/MgO]$_{15}$ multilayers under positive (negative) magnetic fields are shown in Fig. 1C (1D). When a pulse voltage of duration 100 μs and amplitude $V_h = 0.59$ V is applied into the upper heater (H1), the original disordered stripe domains transform into densely packed skyrmions, as shown in Fig. 1C for $\mu_0 H_\perp$ = -25.6 mT and Fig. 1D for $\mu_0 H_\perp$ = 25.4 mT, respectively. Since skyrmion topological charge $Q = {1}/{4\pi} \int \boldsymbol{m} \cdot (\partial_x \boldsymbol{m} \times \partial_y \boldsymbol{m}) dx dy$ is an odd function of the normalized magnetization $\boldsymbol{m}$, it switches its sign under reversed external magnetic fields, as evidenced by the opposite color contrasts in Figs. 1C and 1D. The temperature at the upper/lower edges are $T^S_{U-E} = 399.3$ K and $T^S_{L-E} = 392.5$ K, and the calculated temperature gradient in the multilayer is $\Delta T(x) = 1.6$ K/μm, respectively. The diameter of skyrmion is estimated to be around 140 nm by fitting the skyrmion profile (see Part 3 of Supplementary Materials), and the skyrmion density is 6.8/μm$^2$. Note that similar phenomena shown in Fig. 1C(1D) can also be repeated by using the lower heater (H2) located on the opposite sides of multilayer, as shown in Extended Data S4.

To resolve the detailed intermediate processes of the skyrmion generation, a sequence of smaller pulse voltages was adopted in order to reduce the generation efficiency. At $\mu_0 H_\perp$ = -19.3 mT, stripe domains prevail in the multilayer. Following the increased number of pulses, three distinct behaviors are identified, as shown in Fig. 2A: (1) Nucleation of skyrmions from the hot edge. (2) Transformation from the existing stripe domains into isolated skyrmions. (3) The unidirectional motion of thermally generated skyrmions from the hot region towards the cold region. Intriguingly, under a saturated ferromagnetic (FM) background, coexisting stripe domains and isolated skyrmions can also be generated near the hot edge and from the interior of multilayers (presumably around structural defects for which exhibits a low energy barrier[31]), as shown in Fig. 2B. These characteristics can be clearly seen in Extended Data S1-S3 and in Supplementary Movies (1-5).

Fig. 2C illustrates the thermal generation of skyrmions in the [Pt/Co/Ta]$_{15}$ multilayer from a fully saturated FM background ($\mu_0 H_\perp$ = -47.8 mT). The diameter of skyrmions is around 95 nm in this multilayer, owing to the elevated DMI strength. After applying a pulse voltage (100 μs and $V_h$ = 0.68 V, $T_{U-E}^S$ = 436 K) across the upper heater (H1), skyrmions are solely generated from the hot edge, and then propagate towards the cold edge upon applying the next pulse voltage. Note that the pinning effect of skyrmions in the [Pt/Co/Ta]$_{15}$ is strong, compared with [Ta/Co$_{20}$Fe$_{60}$B$_{20}$/MgO]$_{15}$ and [Pt/Co$_{60}$Fe$_{20}$B$_{20}$/MgO/Ta]$_{15}$ multilayers[10]. The representative thermal generation of skyrmions in the [Pt/Co$_{60}$Fe$_{20}$B$_{20}$/MgO/Ta]$_{15}$ is also provided in Extended Data S5. Since both the pinning effect and the magnetic damping intimately correlate with the structural inhomogeneities or defects, we empirically used the damping parameter ($\alpha$) as an indicator to show the effect of pinning on skyrmion generation. As shown in Fig. 2D, a monotonic increase of threshold temperatures ($T_{U-E}^{th}$) are required to produce densely packed skyrmion phases ($\mu_0 H_\perp$ = 24.6 mT). Since a large damping parameter reduces the thermal activation rate of skyrmion crossing the energy barrier, it raises $T_{U-E}^{th}$. Other factors such as exchange interactions, boundary geometry are also important in determining $T_{U-E}^{th}$, which requires further investigations.

To quantify the skyrmion generation rate as a function of pulse duration and amplitude, the skyrmions were counted. When 375 K < $T_{U-E}^S$ < 385 K (intercept at the *x* axis), the increased $T_{U-E}^S$ results in a linear increase of skyrmion generation rate by 27/K, as shown in Fig. 2E. The intercept at 375 K marks a threshold temperature where substantial skyrmions are generated after applying pulse voltage with duration 100 μs. A phase diagram summarizing the evolution of stripe domains, coexisting stripe domains and skyrmions, densely packed skyrmion lattice and saturated FM states as a function of magnetic field and temperature of the hot edge ($T_{U-E}^S$) is provided in Fig. 2F for the [Ta/Co$_{20}$Fe$_{60}$B$_{20}$/MgO]$_{15}$ multilayer. A similar phase diagram on the duration of pulse voltages ($V_h = 0.53$ V) and $\mu_0 H_\perp$ is also displayed in Extended Data S6. In both phase diagrams, four different magnetic phases can be clearly distinguished. When magnetic fields are small ($|\mu_0 H_\perp| < 10$ mT), increase of $T_{U-E}^S$ only result in a configurational change of stripe domains. This is due to the fact that the skyrmion phase is locally unstable under weak magnetic fields. When $|\mu_0 H_\perp| > 10$ mT, thermal fluctuations in the range of $375$ K $< T_{U-E}^S < 397$ K produce a coexisting phase of stripe domains and sparsely distributed skyrmions. This can be attributed to the stochastic nature of the thermally assisted skyrmion nucleation by overcoming the energy barrier separating the skyrmion and stripe phase. At higher $T_{U-E}^S > 397$ K, thermal fluctuations further promote the nucleation of skyrmions and the system enters to a densely packed skyrmion lattice state. When $|\mu_0 H_\perp| > 45$ mT, the system remains in the FM state in the experimentally accessible temperature range, which indicates that the FM state is the globally stable state in this field region[32].

**Diffusion of skyrmions driven by temperature gradients**

When skyrmion generation is efficient at the hot edge, there exists an increased repulsive force between the newly generated skyrmions and existing skyrmions. This naturally results in a directional motion of skyrmions from the hot region towards the cold region. However, this directional diffusion can be also attributed to the competition among entropic forces[22], magnonic spin torques[17,18,20,21], and thermal SOTs[15]. To separate these mechanisms, we designed a nanowire pointing to the heater. Skyrmion generation at the hot end is minimized due to the tip-like geometry, as shown in Fig. 3A. In fact, the implementation of this type of device may enable a single skyrmion to be generated in a controllable manner. Additionally, the skyrmion generation can be suppressed if one controls the temperature at the tip ($T_{U-E}^S$) to be lower than the threshold temperature for skyrmion generation ($T_{U-E}^{th}$), as guided by the phase diagram shown in Fig. 2F. After applying pulse voltages of duration 500 ms and amplitudes in the range of $0.18 - 0.23$ V (corresponding to $338$ K $< T_{U-E}^S < 365$ K) into the heater, the thermal generation of skyrmion is completely suppressed. The resulting unidirectional diffusion driven by temperature gradients along the nanowire is clearly observed, with $\Delta T(x)$ in the range of $0.35$ K/μm $- 0.56$ K/μm, as shown in Fig. 3A. Note that a few skyrmions were annihilated upon applying a larger $\Delta T(x)$. During the diffusion, sparsely distributed skyrmions were aligned in the center of the [Ta/Co$_{20}$Fe$_{60}$B$_{20}$/MgO]$_{15}$ nanowire, which can be clearly seen from the stochastic trajectory shown in Fig. 3A-8. This occurs as a result of the balanced skyrmion-edge interaction from both edges[33]. Those results clearly indicate a Brownian-like diffusion[14]. By taking the ratio between displacements and pulse durations, the diffusion velocity is also calculated and shown in Fig. 3A-9. The observed nonlinear velocity and the absence of skyrmion Hall effect are consistent with the stochastic nature of skyrmion diffusion. Nevertheless, our experiments clearly show that the thermal diffusion of skyrmions is dominant over the opposite motion driven by magnonic spin torque[17,18,20].

**Theory on the skyrmion generation and dynamics**

All the aforementioned experimental observations can be well addressed in a unified theoretical setting as follows. In our multilayer, there exist competing metastable phases: stripe domains, mixture of stripe domains and skyrmions, skyrmion lattices and saturated FM states. Near the (hot) edge, skyrmions can be generated without meeting any singularity. Together with the twisted edge spins by the unbalanced interfacial DMI[31,34], the energy barriers for skyrmion generation is relatively low. Local heating at the edge could thus efficiently facilitate skyrmion generation from different magnetic phases. Our calculations based on the statistical rate theory[35] and Monte Carlo (MC) simulations given in Supplementary Materials confirm that the edge is the dominant source for skyrmion generation. Note that skyrmions can also be generated from the structural defects in the interior of multilayers, and by coalescing of the stripe domains[36], which also exhibit low energy barriers.

Our experiments can be numerically reproduced by solving the stochastic Landau-Lifshitz-Gilbert equation in the presence of temperature gradients. In our micromagnetic simulations, other sources that could influence the dynamics of skyrmions, including the thermal spin Hall effect, the magnonic spin torques arising from thermally excited magnons[17,18,20], and the repulsive interaction between skyrmions are taken into account. The system Hamiltonian reads as $\mathcal{H} = J_{\text{ex}}/2 \, (\nabla \mathbf{S})^2 + D[S_z(\nabla \cdot \mathbf{S}) - (\mathbf{S} \cdot \nabla)S_z] - \mathbf{H}_a \cdot \mathbf{S}$ with $J_{\text{ex}}$ being the strength of exchange interaction, $D$ the interfacial DMI parameter, and $\mathbf{H}_a$ the perpendicular magnetic fields. The equation of motion for spins (**S**) can be written as:

$$\partial_t \mathbf{S} = -\gamma \mathbf{S} \times (\mathbf{H}_{\text{eff}} + \mathbf{h}_n) + \alpha \mathbf{S} \times \partial_t \mathbf{S}, \qquad (1)$$

where $\gamma$ is the gyromagnetic ratio, $\mathbf{H}_{\text{eff}} = -\delta\mathcal{H}/\delta\mathbf{S}$ is the effective field, $\mathbf{h}_n$ is the random thermal fluctuating field at $T(x) = kx$. The choice of $k$ is to ensure the temperature at the hot edge ($T_{U-E}^S$) is comparable to the energy barrier $\Delta$, such that appreciable numbers of skyrmions can be generated at the hot region in the time scale of simulations. Using the saturated FM state as an initial state, the thermal generation of skyrmions from the hot edge of devices can be seen, which is followed by a unidirectional diffusion towards the cold region, as shown in Fig. 3B. If periodic stripe domains were used as the initial state instead, similar phenomena together with a morphological transition from stripe domains to skyrmions, were identified and shown in Extended Data S7. In both cases, the multilayer is eventually filled with densely packed skyrmions. Once the temperature at the hot side is reduced, skyrmion nucleation is suppressed, and our simulation reproduces thermal diffusion of skyrmions driven by a temperature gradient reported in Fig. 3, as shown in Extended Data S8. Calculations performed by considering materials specific parameters and by taking into account the magnetostatic field added to the effective field of Eq. 1 show qualitatively similar results[37].

By assuming the skyrmion diffusion is much faster than the skyrmion generation rate, the skyrmion-skyrmion repulsive interaction can be neglected. In this case, the thermal diffusion of a skyrmion driven by temperature gradients can be studied (as experimentally demonstrated in Fig. 3A). After establishing the local thermal equilibrium with a linear $\Delta T(x)$, the thermal generation and subsequent diffusion of skyrmions can be described by the Fokker-Planck equation in dimensionless units as follows[17,20]:

$$\partial_t \bar{P} + \partial_x[\bar{F}_m \bar{P}] - G_{xx}\partial_x(T\partial_x \bar{P}) = \omega_s \exp\left[-\frac{\Delta}{k_B T}\right], \qquad (2)$$

where $\bar{P}(x)$ is the probability of finding skyrmions at time $t$ and position $x$. The second term on the left describes the average drift velocity ($\bar{F}_m$) of skyrmion due to the magnonic spin torque, entropic forces and spin torque generated by thermoelectric currents, in competition with pinning force due to defects. The third term on the left describes the diffusion of skyrmions with $G_{xx}$ being the gyrotropic coupling component. The term at the right side of Eq. (2) corresponds to the thermal activation of skyrmions with $\Delta$ being the energy barrier and $\omega_s$ being the attempt frequency. In the regime $k_B T \ll \Delta$, skyrmion nucleation is greatly suppressed. Under small temperature gradient, $\bar{F}_m \bar{P} \gg G_{xx} T \partial_x \bar{P}$. When the pinning is weak, free drift of the existing skyrmions can be enabled, and the direction of motion depends on the relative strength of the magnonic spin torque and SOT of thermoelectric currents.

When $k_B T$ is comparable with $\Delta$, the thermally assisted skyrmion generation becomes important. Subsequently, the dynamics is governed by the diffusion of skyrmions from the hot region towards the cold region. For a clear demonstration, the time evolution of the skyrmion densities for varying amplitudes of $\bar{F}_m = 0.2$ and $\bar{F}_m = 0$ are calculated and shown in Figs. 3C and 3D, respectively. It is evident that skyrmions can be thermally nucleated at the hot edge regardless of the choice of magnonic spin torque and spin torques from thermoelectric current ($\bar{F}_m \geq 0$), and subsequently diffuse to the whole system with a gradient in the density, as indicated by the integrated probability for skyrmion distribution.

**Electrical detection of thermally generated skyrmions**

The presence of skyrmions could significantly affect the dynamics of conduction electrons[2,4]. One well-known example is the anomalous Nernst effect (ANE)[15,38], in which the Nernst voltage reads as $V_{ANE} \propto M_z \cdot \Delta T(x)$, with $M_z$ representing the change of perpendicular magnetization that is related to the change of total number of skyrmions. This motivates the electrical detection of thermally generated skyrmions using the same type of device in Fig. 1A. Note that skyrmions contribute to the Nernst voltage through their magnetization ($M_z$) and its topology ($Q$). The skyrmion diffusion could also produce a topological Nernst voltage, but it is expected to be much smaller[39,40]. By referring to resistance-current ($R$-$I_{\text{Heater}}$) and temperature-resistance ($R$-$T$) curves of both heaters (H1,2), temperatures therein can be experimentally determined (labeled as $T_{1,2}^M$), which are consistent with our COMSOL simulations. The temperature differences between $T_{U-E}^S$ and $T_{L-E}^S$ thus determine the temperature gradients $\Delta T(x)$ across the multilayer, as shown in Figs. 4A and 4B. Under temperature gradients $\Delta T(x) = \pm 2.8$ K/μm generated by either the upper (H1) or the lower (H2) heater, opposite signs of voltages can be observed as shown in Fig. 4C. More ANE data can be found in Extended Data S9.

At fixed $\Delta T(x)$ and constant $B_\perp$, the change of $V_{ANE}$ reflects the change of magnetization and hence the total skyrmion numbers, as given in Fig. 4C. To demonstrate the electrical detection of a single skyrmion, we first applied a large current $I_{\text{Heater}} = 6$ mA ($T_{U-E}^S = 428$ K $> T_{U-E}^{\text{th}}$) to generate skyrmions and kick off their diffusion. $V_{ANE}$ was then measured under a smaller current $I_{\text{Heater}} = 3$ mA to maintain $\Delta T(x) = 0.9$ K/μm, where no new skyrmions were generated during the measurement [($T_{U-E}^S$(325 K) $< T_{U-E}^{\text{th}}$(375 K)]. The time dependent $V_{ANE}$ measured at $\mu_0 H_\perp = 25$ mT and the corresponding current pulse profile are displayed in Fig. 4D. Right after switching to current $I_{\text{Heater}}$= 3 mA, $V_{ANE}$ decreases sharply, signifying the reduction of total skyrmion number and/or change in magnetization distribution. Interestingly, we observed many discretized steps in the time evolution of $V_{ANE}$ where the change of $V_{ANE}$ is discretized by $\Delta V_{ANE} = 90 \pm 10$ nV. Each discretized jump can be naturally explained by the annihilation of a single skyrmion in the multilayer, which results in change of $M_z$ when approaches the thermal equilibrium[41]. Compared to our micromagnetic simulations given in Extended Data S8 and diffusion dynamics in Fig. 3A, we conclude that these discretized jumps correspond to the disappearance of a single skyrmion, likely through sample edges or structural defects in the interior of the device. Our experiments thus confirm that anomalous Nernst effect can be used for electrical detection of single skyrmion in similar multilayers[42].

**Perspective**

We have demonstrated the on-chip thermal generation, manipulation and detection of nanoscale skyrmions in metallic multilayers. Specifically, a unidirectional diffusion of skyrmions from the hot region toward the cold region is observed which can be attributed to combined contribution from repulsive forces between skyrmions, entropic forces, magnonic spin torque and thermal spin-orbit torques. More importantly, discretized jumps in anomalous Nernst voltages were measured for electrical detection of a single skyrmion. Our experiments could serve as a complementary approach that can be integrated with the existing electrical manipulation scheme, and can be extended for studying different types of topological spin textures in various ferromagnetic and ferroelectric materials[27,43-45]. In particular, insulating skyrmion-hosting materials where electrical currents are inapplicable, are perfect testing platforms[26,27]. Based on the Onsager reciprocal relation, the interacting thermal currents and skyrmion lattices could also result in many topological phenomena to be examined, including the topological magnon Hall effect, and thermally induced skyrmion Hall effect, just to name a few examples[4,23]. Thus, the thermal generation, manipulation and detection of skyrmions could largely expand the current paradigm of spintronics, and may lead to future discoveries in skyrmionics, magnonics and spin caloritronics.

**Methods**

Interfacially asymmetric multilayers [Ta(30Å)/Co$_{20}$Fe$_{60}$B$_{20}$(11Å)/MgO(20Å)]$_{15}$, [Pt(20Å)/Co$_{60}$Fe$_{20}$B$_{20}$(11Å)/MgO(14Å)/ Ta(10Å)]$_{15}$ and [Pt(15Å)/Co(10Å)/Ta(5Å)]$_{15}$ were grown onto semi-insulating Si substrates covered with 300 nm thick thermally formed SiO$_2$ layer, and onto a 100 nm thick Si$_3$N$_4$ membrane on top of Si supporting frames. These films were made by using a dc magnetron sputtering system (AJA, Orion 8) at room temperature under Ar pressure 3 mTorr with a base pressure of the sputtering chamber $< 2 \times 10^{-8}$ Torr. The Si$_3$N$_4$ membranes we used in this study are from YW MEMS (Suzhou). Co., Ltd. Multilayer channels on 100 nm thick Si$_3$N$_4$ membranes were first patterned by using electron beam lithography and followed by a lift-off process, which were annealed in vacuum for 30 minutes to induce the perpendicular magnetic anisotropy. Subsequently, Ta (20 nm)/Pt (50 nm) electrodes were deposited. A Quantum Design superconducting quantum interference device (SQUID) magnetometer was used to measure the magnetic properties. Damping parameters in multilayers were determined from ferromagnetic resonance (FMR). In order to probe the dominant out-of-plane X-ray magnetic circular dichroism (XMCD) contrast, samples on 100 nm thick Si$_3$N$_4$ membrane for XMCD imaging were positioned with the plane normal to the incident circularly polarized X-ray beam. The best magnetic contrast is obtained at Fe $L_3$ edge 708.5 eV for the [Ta/Co$_{20}$Fe$_{60}$B$_{20}$/MgO]$_{15}$ multilayers. Due to the less (absent) content of Fe and the accompanied weak signal, the magnetic contrasts for the [Pt/Co$_{60}$Fe$_{20}$B$_{20}$/MgO/Ta]$_{15}$ and [Pt(15Å)/Co(10Å)/Ta(5Å)]$_{15}$ multilayers were obtained at Co $L_3$ edge 778.5 eV. Voltage pulses supplied into the on-chip heaters were provided by an Agilent 81150A arbitrary waveform generator and monitored with a 50 Ω terminated real-time oscilloscope, through which the current flowing in the heater can be calculated. Note that for anomalous Nernst measurement, the length of the multilayer channel is 30 μm.

Calibration of temperatures of both heaters were made by firstly measuring the resistance change of heaters as a function of current/voltage, through comparing with temperature-dependent resistance change, temperatures at both heaters can be identified. The presence of temperature gradients is further confirmed through anomalous Nernst measurement. The presence of approximately linear temperature gradients in the sample is also experimentally verified, and discussed in Part 8 of Supplementary Materials. Temperature profiles of the devices were simulated by using COMSOL Multiphysics software. We implemented a conjoint *Joule Heating* module, which includes the *AC/DC* module to apply the pulse voltages. The *Heat Transfer* module was used to describe the heat flow and temperature distribution. Since the most part of the device is located on top of the 500 μm × 500 μm $Si_3N_4$ (100 nm), we therefore limited our simulation in this area. For heaters and electrodes, the materials parameters for Pt are used: thickness of 70 nm and an electrical conductivity of $8.9 \times 10^6$ S/m. The multilayer channels were simplified as: $[Ta(15nm)/Co_{20}Fe_{60}B_{20}(5.5nm)/MgO(10nm)]_3$, $[Pt(10nm)/Co_{60}Fe_{20}B_{20}(5.5nm)/MgO(9nm)/Ta(5nm)]_3$ and $[Pt(7.5nm)/Co(5nm)/Ta(2.5nm)]_3$, to save the simulation time. All parameters including density $\rho$, specific heat capacity $C_\rho$, thermal conductivity $\kappa$, and electrical conductivity $\sigma$ used in COMSOL simulations are shown in Table S1 of Supplementary Materials. We used the fixed boundary condition with the temperature fixed at 293 K. Micromagnetic simulation studies were independently carried out using a home-built code and by using a state-of-the-art micromagnetic solver, PETASPIN, full description which can be found in Supplementary Materials[37].

**Acknowledgements**

Work carried out at Tsinghua was supported by the Basic Science Center Project of NSFC (Grant No. 51788104), the National Key R&D Program of China (Grant Nos. 2017YFA0206200 and 2016YFA0302300), the National Natural Science Foundation of China (Grant No. 11774194, 51831005, 1181101082, 11804182), Beijing Natural Science Foundation (Grant No. Z190009), Tsinghua University Initiative Scientific Research Program and the Beijing Advanced Innovation Center for Future Chip (ICFC). The work at LANL was carried out under the auspices of the U.S. DOE NNSA under contract No. 89233218CNA000001 through the LDRD Program, and was supported by the Center for Nonlinear Studies at LANL. Works at the ALS were supported by U.S. Department of Energy (DE-AC02-05CH11231). M.-Y. Im acknowledges support from the National Research Foundation (NRF) of Korea funded by the Ministry of Education, Science and ICT (2018K1A4A3A03075584, 2016M3D1A1027831), DGIST R&D program of the Ministry of Science, ICT and future Planning (18-BT-02) and support by Lawrence Berkeley National Laboratory through the Laboratory Directed Research and Development (LDRD) Program. R.T. and G.F. thank the project "ThunderSKY" funded from the Hellenic Foundation for Research and Innovation (HFRI) and the General Secretariat for Research and Technology (GSRT) under Grant No. 871. Authors wish to thank Naoto Nagaosa, Markus Garst, Jiadong Zang, Xichao Zhang, Guoqiang Yu and Yayu Wang for fruitful discussions.

## Author contributions

W.J. conceived the idea and designed the experiments. H.Z. and T.X. fabricated the thin film. Z.W., H.Z., L.Z., Y.D. and C.S. performed lithographic processing. Z.W. did the COMSOL simulation. M.G., H.B. and H.W. did the thermoelectric measurements. Y.Y., H.Z. and W.H. carried out ferromagnetic resonance experiments. S.L. performed atomistic micromagnetic simulation and Fokker-Planck calculation. R.T. and G. F. performed the layer dependent micromagnetic simulations. M.C and G.F. implemented the micromagnetic solver for multilayer calculations. Z.W., S.J., H.H., K.L., S.L., W.C., M.I and W.J. performed the full field soft X-ray microscope imaging experiments and data analysis. W.J. and S.L. wrote the manuscript with inputs from all authors.

## Additional information.

Supplementary information is available in the online version of the paper. Correspondence and requests for materials should be addressed to S.L. and W.J.

## Competing financial interests

Authors declare no competing financial interests.

## References

1    Yu, X. Z. *et al.* Real-space observation of a two-dimensional skyrmion crystal. *Nature* **465**, 901-904 (2010).
2    Jonietz, F. *et al.* Spin transfer torques in MnSi at ultralow current densities. *Science* **330**, 1648-1651 (2010).
3    Fert, A., Cros, V. & Sampaio, J. Skyrmions on the track. *Nature Nanotechnology* **8**, 152-156 (2013).
4    Nagaosa, N. & Tokura, Y. Topological properties and dynamics of magnetic skyrmions. *Nature Nanotechnology* **8**, 899-911 (2013).
5    Bogdanov, A. N. & Rößler, U. K. Chiral symmetry breaking in magnetic thin films and multilayers. *Physical Review Letters* **87**, 037203 (2001).
6    Heinze, S. *et al.* Spontaneous atomic-scale magnetic skyrmion lattice in two dimensions. *Nature Physics* **7**, 713-718 (2011).
7    Romming, N. *et al.* Writing and deleting single magnetic skyrmions. *Science* **341**, 636-639 (2013).
8    Jiang, W. *et al.* Blowing magnetic skyrmion bubbles. *Science* **349**, 283-286 (2015).
9    Chen, G., Mascaraque, A., N'Diaye, A. T. & Schmid, A. K. Room temperature skyrmion ground state stabilized through interlayer exchange coupling. *Appl Phys Lett* **106**, 242404 (2015).
10   Woo, S. *et al.* Observation of room-temperature magnetic skyrmions and their current-driven dynamics in ultrathin metallic ferromagnets. *Nature Materials* **15**, 501-506 (2016).

11  Moreau-Luchaire, C. *et al.* Additive interfacial chiral interaction in multilayers for stabilization of small individual skyrmions at room temperature. *Nature Nanotechnology* **11**, 444-448 (2016).

12  Fert, A., Reyren, N. & Cros, V. Magnetic skyrmions: advances in physics and potential applications. *Nat Rev Mater* **2**, 17031 (2017).

13  Everschor-Sitte, K., Masell, J., Reeve, R. M. & Klaui, M. Perspective: Magnetic skyrmions-Overview of recent progress in an active research field. *Journal of Applied Physics* **124**, 240901 (2018).

14  Jakub Zázvorka *et al.* Thermal skyrmion diffusion used in a reshuffler device. *Nature Nanotechnology* **14**, 658-661 (2019).

15  Bauer, G. E. W., Saitoh, E. & van Wees, B. J. Spin caloritronics. *Nature Materials* **11**, 391-399 (2012).

16  Koshibae, W. & Nagaosa, N. Creation of skyrmions and antiskyrmions by local heating. *Nature Communications* **5**, 5148 (2014).

17  Lin, S. Z., Batista, C. D., Reichhardt, C. & Saxena, A. ac Current Generation in Chiral Magnetic Insulators and Skyrmion Motion induced by the Spin Seebeck Effect. *Physical Review Letters* **112**, 187203 (2014).

18  Kovalev, A. A. & Tserkovnyak, Y. Thermoelectric spin transfer in textured magnets. *Physical Review B* **80**, 100408 (2009).

19  Everschor, K. *et al.* Rotating skyrmion lattices by spin torques and field or temperature gradients. *Physical Review B* **86**, 054432 (2012).

20  Kong, L. Y. & Zang, J. D. Dynamics of an Insulating Skyrmion under a Temperature Gradient. *Physical Review Letters* **111**, 067203 (2013).

21  Mochizuki, M. *et al.* Thermally driven ratchet motion of a skyrmion microcrystal and topological magnon Hall effect. *Nature Materials* **13**, 241-246 (2014).

22  Wild, J. *et al.* Entropy-limited topological protection of skyrmions. *Science Advances* **3**, e1701704 (2017).

23  Shiomi, Y., Kanazawa, N., Shibata, K., Onose, Y. & Tokura, Y. Topological Nernst effect in a three-dimensional skyrmion-lattice phase. *Physical Review B* **88**, 064409 (2013).

24  Chumak, A. V., Vasyuchka, V. I., Serga, A. A. & Hillebrands, B. Magnon spintronics. *Nature Physics* **11**, 453-461 (2015).

25  R. Tomasello *et al.* A strategy for the design of skyrmion racetrack memories. *Scientific Reports* **4**, 6784 (2014).

26  Seki, S., Yu, X. Z., Ishiwata, S. & Tokura, Y. Observation of Skyrmions in a Multiferroic Material. *Science* **336**, 198-201 (2012).

27  Kézsmárki, I. *et al.* Néel-type skyrmion lattice with confined orientation in the polar magnetic semiconductor $GaV_4S_8$. *Nature Materials* **14**, 1116 (2015).

28  Lemesh, I. *et al.* Current-Induced Skyrmion Generation through Morphological Thermal Transitions in Chiral Ferromagnetic Heterostructures. *Advanced Materials* **30**, 1805461 (2018).

29  Legrand, W. *et al.* Hybrid chiral domain walls and skyrmions in magnetic multilayers. *Science Advances* **4**, eeat0415 (2018).

30  Li, W. *et al.* Anatomy of Skyrmionic Textures in Magnetic Multilayers. *Adv Mater* **31**, 1807683 (2019).

31  Muller, J., Rosch, A. & Garst, M. Edge instabilities and skyrmion creation in magnetic layers. *New Journal of Physics* **18**, 065006 (2016).

32  Bottcher, M., Heinze, S., Egorov, S., Sinova, J. & Dupe, B. B-T phase diagram of Pd/Fe/Ir(111) computed with parallel tempering Monte Carlo. *New Journal of Physics* **20**, 103014 (2018).

33   Zhang, X. C. *et al.* Skyrmion-skyrmion and skyrmion-edge repulsions in skyrmion-based racetrack memory. *Scientific Reports* **5**, 7643 (2015).
34   Rohart, S. & Thiaville, A. Skyrmion confinement in ultrathin film nanostructures in the presence of Dzyaloshinskii-Moriya interaction. *Physical Review B* **88**, 184422 (2013).
35   Bessarab, P. F. *et al.* Lifetime of racetrack skyrmions. *Scientific Reports* **8**, 3433 (2018).
36   Lin, S.-Z. Edge instability in a chiral stripe domain under an electric current and skyrmion generation. *Physical Review B* **94**, 020402(R) (2016).
37   Tomasello, R. *et al.* Micromagnetic understanding of the skyrmion Hall angle current dependence in perpendicularly magnetized ferromagnets. *Physical Review B* **98**, 224418 (2018).
38   Kim, D. J. *et al.* Observation of transverse spin Nernst magnetoresistance induced by thermal spin current in ferromagnet/non-magnet bilayers. *Nature Communications* **9**, 1400 (2018).
39   Zeissler, K. *et al.* Discrete Hall resistivity contribution from Neel skyrmions in multilayer nanodiscs. *Nature Nanotechnology* **13**, 1161-1166 (2018).
40   Maccariello, D. *et al.* Electrical detection of single magnetic skyrmions in metallic multilayers at room temperature. *Nature Nanotechnology* **13**, 233-237 (2018).
41   Kanazawa, N. *et al.* Discretized topological Hall effect emerging from skyrmions in constricted geometry. *Physical Review B* **91**, 041122 (2015).
42   Scarioni, A. F. *et al.*   (Preprint at: https://arxiv.org/pdf/2001.10251., 2020).
43   Nayak, A. K. *et al.* Magnetic antiskyrmions above room temperature in tetragonal Heusler materials. *Nature* **548**, 561 (2017).
44   Yu, X. Z. *et al.* Transformation between meron and skyrmion topological spin textures in a chiral magnet. *Nature* **564**, 95-98 (2018).
45   Das, S. *et al.* Observation of room-temperature polar skyrmions. *Nature* **568**, 368-372 (2019).

**Figure Captions**

**Figure 1. Thermal generation of skyrmions via using the on-chip heaters.** Fig. A is an optical image of the integrated device with two identical Ta/Pt heaters on top of a 100 nm thick $Si_3N_4$ membrane. The imaging area is marked as a blue circle in the center of the $[Ta/CoFeB/MgO]_{15}$ multilayer channel. The computed temperature profile is shown in the left upper panel of Fig. B with a pulse voltage of amplitude $V_h = 0.59$ V and duration 100 μs. Shown in the left lower panel of Fig. B is a linescan of temperature profile (cyan line) from which a linear temperature gradient in the multilayer can be found. Temperatures of both heaters ($T_{H1,2}^S$) and the upper and the lower edges of the multilayer channel were computed and labeled as $T_{U-E}^S$ and $T_{L-E}^S$, respectively. Shown in Figs. C and D are the transformation from the original stripe domains into densely packed skyrmions after applying a pulse voltage of duration 100 μs and amplitude $V_h = 0.59$ V to the upper heater (H1) for negative magnetic field ($\mu_0 H_\perp = -25.6$ mT) and positive field ($\mu_0 H_\perp = 25.4$ mT), respectively. White color corresponds to magnetization downward while black color denotes magnetization upward, respectively.

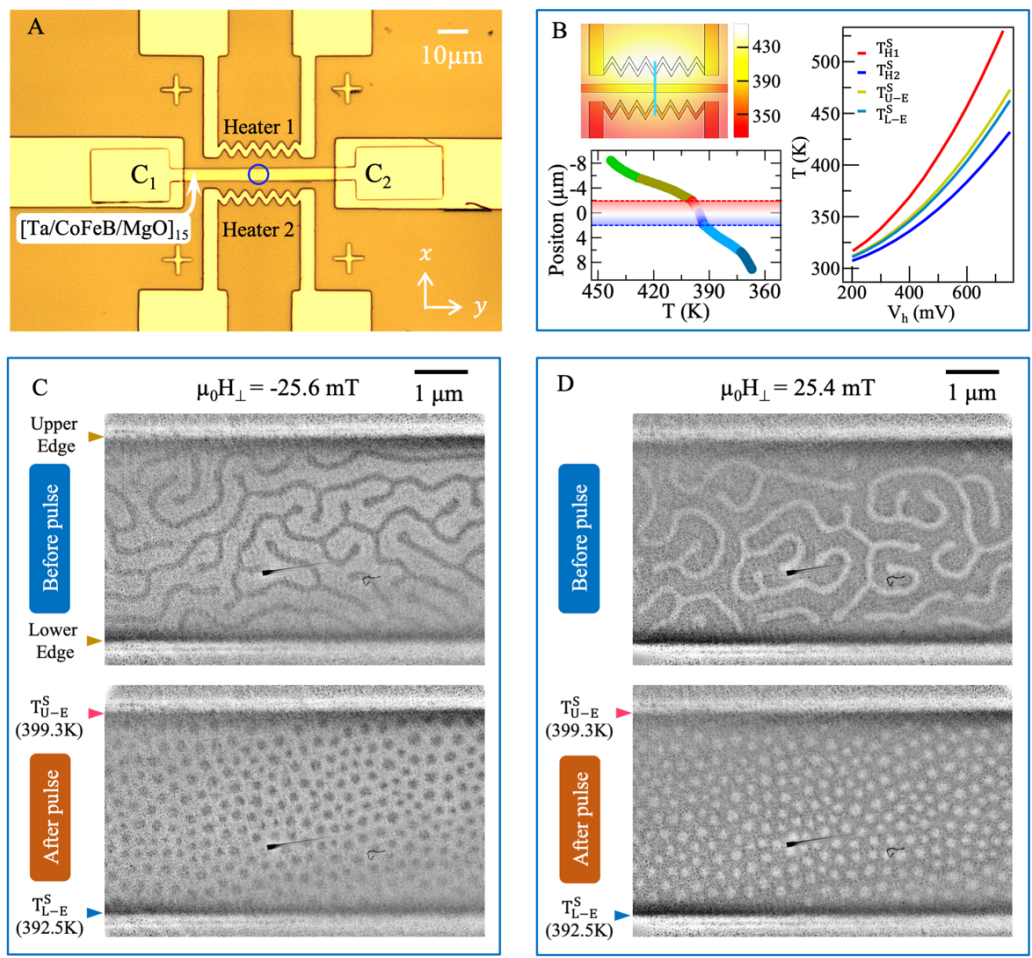

**Figure 2. Transformational dynamics and phase diagram in different multilayers.** Shown in Figs. A and B are consecutive images acquired in the [Ta/CoFeB/MgO]$_{15}$ multilayer at $\mu_0 H_\perp = -19.3$ mT and $\mu_0 H_\perp = -27.6$ mT, before and after applying pulse voltages (duration fixed at 100 μs) into the upper heater (H1), in which the computed temperatures at the hot side ($T_{U-E}^S$) can also be found. Shown in Fig. A are images taken before and after applying the following pulse voltages into the heater: $V_h = 0.514$ V(1$^{st}$), $V_h = 0.517$ V(2$^{nd}$), $V_h = 0.525$ V(3$^{rd}$) and $V_h = 0.531$ V(4$^{th}$) at $\mu_0 H_\perp = -19.3$ mT. Shown in Fig. B are images taken after applying $V_h = 0.556$ V(1$^{st}$), $V_h = 0.560$ V(2$^{nd}$), $V_h = 0.571$ V(3$^{rd}$) and $V_h = 0.586$ V(4$^{th}$) at $\mu_0 H_\perp = -27.6$ mT. While domains are absent from the original image, skyrmions and stripe domains can also be generated. In the [Pt/Co/Ta]$_{15}$ multilayer, skyrmions can be similarly generated from the hot edge (436 K $< T_{U-E}^S <$ 464 K), which then propagate from the hot side towards the cold side followed by the growing area of skyrmion lattices, as shown in Fig. C. The amplitude of voltages (100 μs) at the hot edges are: $V_h = 0.682$ V(1$^{st}$), $V_h = 0.701$ V(2$^{nd}$), $V_h = 0.712$ V(3$^{rd}$) and $V_h = 0.745$ V(4$^{th}$). Shown in Fig. 2D is the dependence of the threshold skyrmion generation temperatures ($T_{U-E}^{th}$) on the damping parameters ($\alpha$) at $\mu_0 H_\perp = 25.4$ mT. Fig. E corresponds to the skyrmion generation rate as a function of $T_{U-E}^S$ for the [Ta/CoFeB/MgO]$_{15}$ multilayer. A phase diagram summarizing the evolution of different magnetic phases as a function of $T_{U-E}^S$ and $\mu_0 H_\perp$ is constructed in Fig. F for the [Ta/CoFeB/MgO]$_{15}$ multilayer.

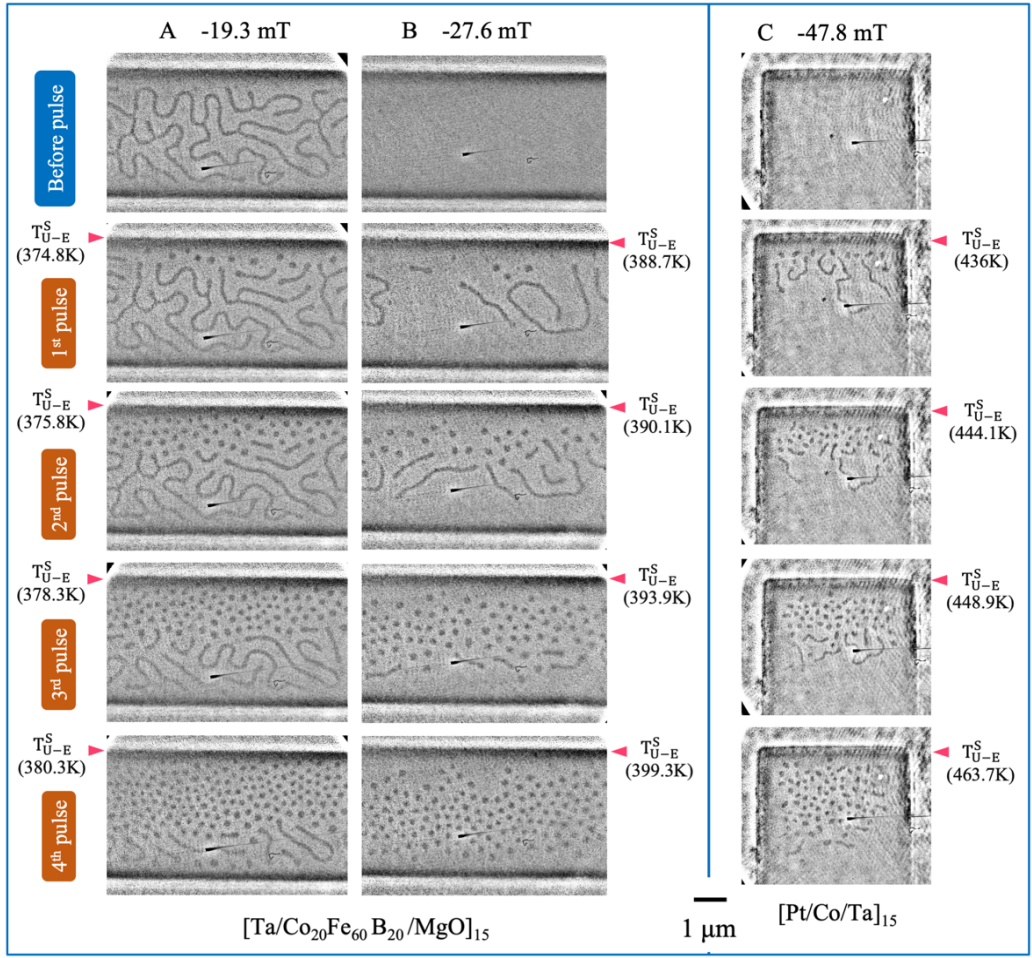
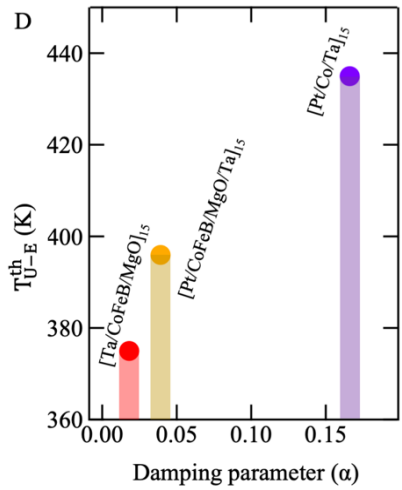
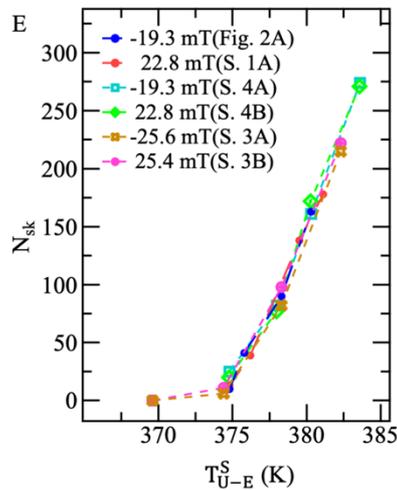
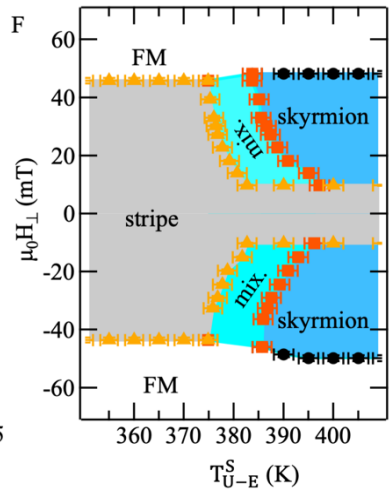

**Figure 3. Experimental demonstration of thermally induced skyrmion diffusion and analytical understandings of skyrmion generation and diffusion.** Show in Fig. A-1 is a scanning electron microscope image. Following the increased temperature gradients (duration is fixed at 500 ms), a unidirectional diffusion of skyrmions from the hot region towards the cold region is observed in a device with a sharp tip in the [Ta/CoFeB/MgO]$_{15}$ multilayer, as shown in Figs. A2-A7. Summarized in Figs. A-8 and A-9 are the diffusion trajectory and the velocity of the selected skyrmions. Marked in the purple hexagon is a pinned skyrmion. Micromagnetic simulation results are shown in Fig. B. Following the increasing number of frames (P$_0$, P$_6$, P$_{23}$, P$_{53}$, P$_{98}$), skyrmions are first nucleated at the hot edge, followed by a directional motion from the hot region towards the cold region. The time lapse between consecutive frames is $\Delta t = 120 J_{ex}/\gamma D^2$. The coefficient of temperature gradient is $k = 0.01 J_{ex}/k_B$. The film thickness is $0.1 J_{ex}/D$. The scale bar is $20 J_{ex}/D$. The integrated skyrmion probability ($\bar{P}$) distribution as a function of position ($J_{ex}/D$) at different time ($t$) obtained by solving the Fokker-Planck equation are shown, Fig. C for a drift velocity $\bar{F}_m = 0.2$ and Fig. D $\bar{F}_m = 0$, respectively. In the calculations, we use the absorbing boundary condition at $x = 0$ by setting $\bar{P}(x = 0) = 0$ at the cold side and $\partial_x \bar{P}(x = L_x) = 0$ at the hot side. The length is in unit of $J_{ex}/D$ and time is in unit of by $4\pi d/\gamma D$.

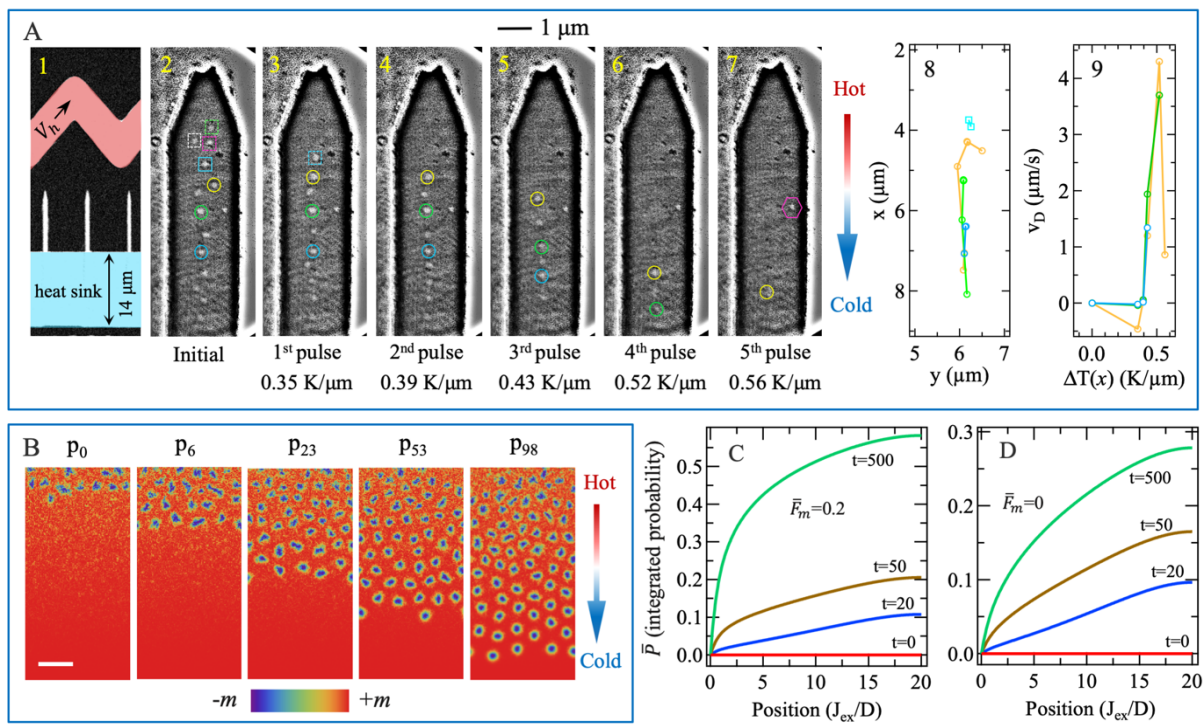

**Figure 4. Electrical detection of skyrmions via the anomalous Nernst effect.** Combined with COMSOL simulations and experimental measurements, temperatures and the accompanied temperature gradients $\Delta T(x)$ in the device are determined and shown in Figs. A and B, respectively. The simulated (line)/measured (dot) temperatures of both heaters are accordingly labeled as $T_{1,2}^S/T_{1,2}^M$. The opposite anomalous Nernst voltages ($V_{ANE}$) measured with the opposite temperature gradients $[\Delta T(x) = \pm 2.8 \text{ K}/\mu\text{m}]$ are shown in Fig. C. At $\mu_0 H_\perp = 25$ mT, after switching off $I_{\text{Heater}} = 6$ mA, a smaller current $I_{\text{Heater}} = 3$ mA that generates $\Delta T(x) = 0.9$ K/μm is applied to measure the time evolution of $V_{ANE}$, in which several "quantized" jumps with $\Delta V_{ANE} = 90 \pm 10$ nV can be found. After saturating the sample above $\mu_0 H_\perp = 500$ mT and reducing the field back to $\mu_0 H_\perp = 25$ mT, $V_{ANE}$ falls back to the same value as before applying $I_{\text{Heater}} = 6$ mA in which the "quantized" jumps are absent.

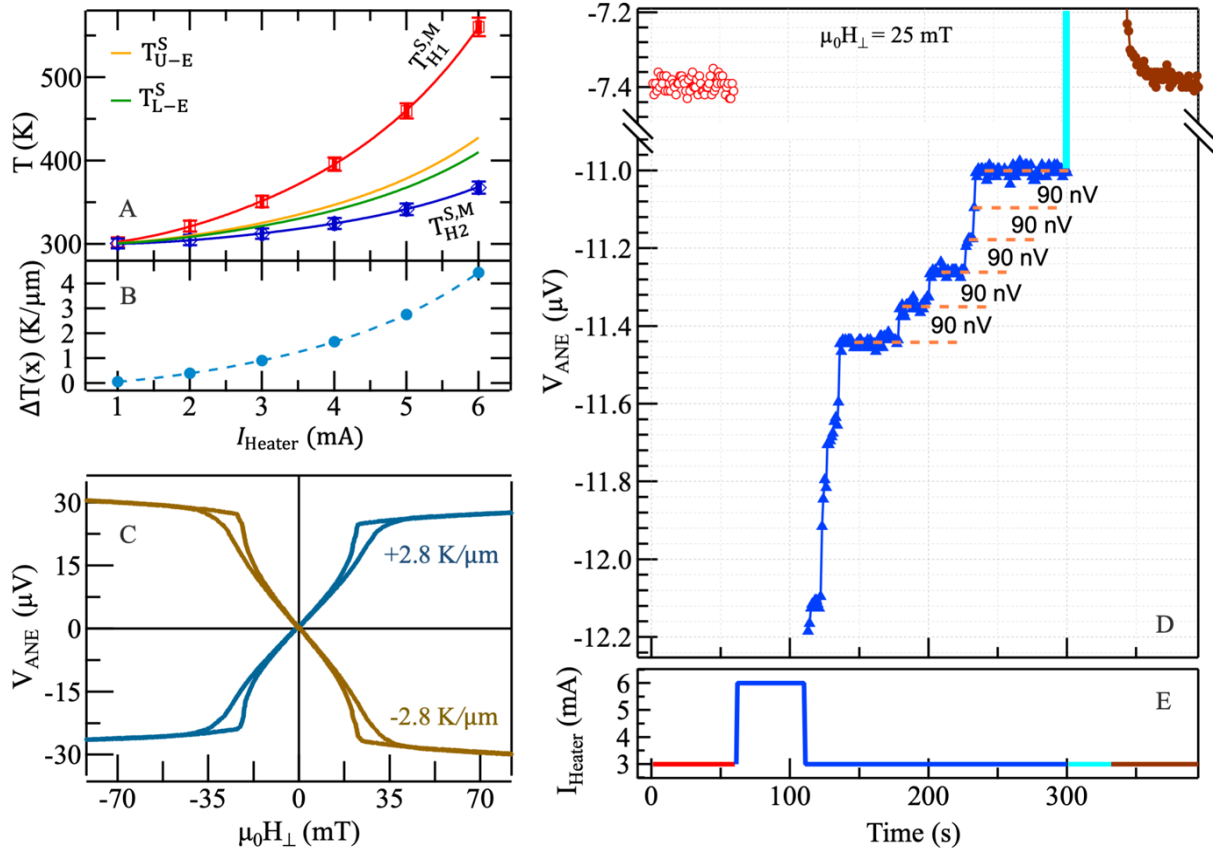